\documentclass[12pt]{article}
\pdfoutput=1

\DeclareFontFamily{OT1}{pzc}{}
\DeclareFontShape{OT1}{pzc}{m}{it}{<-> s * [1.10] pzcmi7t}{}
\DeclareMathAlphabet{\mathpzc}{OT1}{pzc}{m}{it}

\newcommand{\wt}[1]{\widetilde{#1}}

\newcommand{\swisswatch}[1][.3em]{%
	\tikz[baseline=-0.5ex, x=#1, y=#1]{%
		\coordinate (C) at (0,0);     
		\def\Rcase{1.0}               
		\def\Rdial{0.78}              

		\fill[gray!35!black!20!white, rounded corners=0.35]
		(-0.55,1.05) rectangle (0.55,1.75);
		\fill[gray!35!black!20!white, rounded corners=0.35]
		(-0.55,-1.75) rectangle (0.55,-1.05);

		\shade[inner color=gray!5!white, outer color=gray!45!black!20!white]
		(C) circle (\Rcase);

		\fill[gray!50!black!30!white, rounded corners=0.25]
		(\Rcase+0.10,-0.20) rectangle (\Rcase+0.45,0.20);

		\fill[white] (C) circle (\Rdial);

		\foreach \a in {0,6,...,354} {
			\draw[gray!50, line cap=round, line width=0.018]
			(\a:\Rdial-0.06) -- (\a:\Rdial-0.01);
		}

		\foreach \a in {0,30,...,330} {
			\draw[gray!30!black, line cap=round, line width=0.05]
			(\a:\Rdial-0.16) -- (\a:\Rdial-0.02);
		}

		\draw[gray!30!black, line cap=round, line width=0.09]
		(C) -- ++(140:0.42);
		\draw[gray!20!black, line cap=round, line width=0.07]
		(C) -- ++(20:0.65);
		\draw[red!80!black, line cap=round, line width=0.035]
		(C) -- ++(285:0.72);

		\fill[gray!20!black] (C) circle (0.07);
		\fill[white, opacity=0.7] (0.04,0.04) circle (0.04);

		\begin{scope}
			\clip (C) circle (\Rdial);
			\begin{scope}[shift={(0,0.42)}]
				\fill[red!85!black] (-0.16,-0.09) rectangle (0.16,0.09);
				\fill[white]
				(-0.07,-0.03) rectangle (0.07,0.03); 
				\fill[white]
				(-0.03,-0.07) rectangle (0.03,0.07); 
			\end{scope}
		\end{scope}

		\draw[line width=0.03em, draw=gray!40!black!60!white]
		(C) circle (\Rcase);
	}%
}

\usepackage{stmaryrd}

\usepackage{draft}
\usepackage[weather]{ifsym}

\usepackage{hyperref}
\usepackage{graphicx,color,subfig}
\usepackage{cite}
\usepackage{mciteplus}
\usepackage{skak}
\usepackage{empheq}
\usepackage{tikz}
\usepackage{bbm}

\DeclareFontFamily{OT1}{pzc}{}
\DeclareFontShape{OT1}{pzc}{m}{it}{<-> s * [1.10] pzcmi7t}{}
\DeclareMathAlphabet{\mathpzc}{OT1}{pzc}{m}{it}

\usetikzlibrary{calc}
\usetikzlibrary{snakes}
\usetikzlibrary{arrows.meta}
\usetikzlibrary{decorations.pathmorphing}
\usetikzlibrary{decorations.markings}
\usetikzlibrary{bending}
\tikzset{snake it/.style={decorate, decoration=snake}}
\usetikzlibrary{shapes.misc}
\tikzset{cross/.style={cross out, draw=black, minimum size=2*(#1-\pgflinewidth), inner sep=0pt, outer sep=0pt},
cross/.default={1pt}}

\usepackage[T1]{fontenc}
\usepackage{esint}
\usepackage{lmodern}

\def\be#1\ee{\begin{align}#1\end{align}}

\definecolor{dark green}{rgb}{0.7,1,0.64}

\usepackage{listings}
\usepackage{xcolor}

\definecolor{codegreen}{rgb}{0,0.6,0}
\definecolor{codegray}{rgb}{0.5,0.5,0.5}
\definecolor{codepurple}{rgb}{0.58,0,0.82}
\definecolor{backcolour}{rgb}{0.95,0.95,0.92}

\lstdefinestyle{myStyle}{
    belowcaptionskip=1\baselineskip,
    breaklines=true,
    frame=none,
    numbers=none,
    basicstyle=\footnotesize\ttfamily,
    keywordstyle=\bfseries\color{green!40!black},
    commentstyle=\itshape\color{purple!40!black},
    identifierstyle=\color{blue},
    backgroundcolor=\color{gray!10!white},
    tabsize=2,
}

\usepackage{array}

\begin{document}

\unitlength = .8mm

\begin{titlepage}

\begin{center}

\hfill \\
\hfill \\
\vskip 1cm

\title{On the Flux Sectors of Matrix String Theory}

\author{Minjae Cho$^{\text{\Snow}}$, Barak Gabai$^{\text{\swisswatch}}$, Jaroslav Scheinpflug${}^{\text{\FilledWeakRainCloud}}$, Xi Yin${}^{\text{\FilledWeakRainCloud}}$}

\address{$^{\text{\Snow}}$Leinweber Institute for Theoretical Physics, University of Chicago, Chicago, IL 60637 USA
			\\
		$^{\text{\swisswatch}}$Laboratory for Theoretical Fundamental Physics,
		EPFL, \\ Rte de la Sorge, CH-1015, Lausanne \\
		${}^{\text{\FilledWeakRainCloud}}$Jefferson Physical Laboratory, Harvard University,
Cambridge, MA 02138 USA
}

\email{cho7@uchicago.edu, barak.gabai@epfl.ch, jscheinpflug@g.harvard.edu, xiyin@fas.harvard.edu}

\end{center}

\abstract{We analyze the gapped flux vacua of 2D $(8,8)$ $SU(N)$ super-Yang-Mills theory. Based on the matrix string theory duality, we conjecture the spectrum of massive resonances in the gauge theory at large $N$ and beyond the 't Hooft scaling regime.
}

\vfill

\end{titlepage}

\eject

\begingroup
\hypersetup{linkcolor=black}


\endgroup

The matrix string theory (MST) duality \cite{Motl:1997th, Banks:1996my, Dijkgraaf:1997vv} is a proposed duality between the S-matrix of type IIA string theory in ten-dimensional Minkowskian spacetime and the target-space S-matrix of the two-dimensional maximally supersymmetric $U(N)$ gauge theory (SYM) compactified on a circle, and may be viewed as a generalization of the BFSS duality \cite{Banks:1996vh, Seiberg:1997ad, Sen:1997we, MaldaTalk}. Highly nontrivial evidence for the proposal, based on perturbative string amplitudes on one side and conformal perturbation theory of the infrared limit of the gauge theory on the other, has been provided in \cite{Arutyunov:1997gi, Dijkgraaf:2003nw}. 
In this paper, we take seriously a non-perturbative version of the MST duality conjecture and explore its implications for the dynamics of the two-dimensional gauge theory.

\section{The matrix string theory duality}
\label{sec:mstsetup}

The MST duality may be motivated starting from the decoupling limit of the extremal black 1-brane solution of type IIB string theory, described in terms of the string frame metric, the dilaton $\Phi$, and the RR 2-form potential $C_2$ as
\ie\label{dsaiibdoness}
& ds_{\rm str}^2 = (\wt f_1(r))^{-{1\over 2}} (-dt^2 + dx^2) + (\wt f_1(r))^{1\over 2} (dr^2 + r^2 d\Omega_7^2),
\\
& e^\Phi = (\wt f_1(r))^{1\over 2},~~~~ C_2 =  \wt f_1^{-1} dt \wedge dx,
\\
& \wt f_1(r) = {c_1 N\over r^6},~~~~ c_1 = 32 \pi^2 g_B \ell_B^6 = 32\pi^2 g_B^{-{1\over 2}}M_{\rm pl}^{-6},
\fe
where $\ell_B$ is the type IIB string length, and $g_B$ is the type IIB string coupling (defined as the ratio between the F1 and D1 string tensions in the absence of RR axion and dilaton expectation value).
The standard holographic dictionary suggests an exact dual description in terms of the 2D ${\cal N}=(8,8)$ SYM characterized by the $U(N)$ gauge field $A_\mu$, adjoint scalar fields $\phi^i$, and adjoint fermions $\lambda_{\A +}$, $\lambda_{\da -}$. Here $i=1,\cdots,8$ is a vector index with respect to the $so(8)_R$ symmetry, and $\A, \da$ are chiral and anti-chiral spinor indices with respect to the $so(8)_R$. The action reads
\ie
S & = {1\over g_{\rm YM}^2} \int d^2x\, {\rm tr} \bigg( - {1\over 4} F_{\mu\nu} F^{\mu\nu} - {1\over 2} D_\mu \phi^i D^\mu \phi^i + {1\over 4} [\phi^i, \phi^j]^2
\\
&~~~~~~~~~~~~~~~~~~~ -  \lambda_{\A+} D_- \lambda_{\A+} -  \lambda_{\da -} D_+ \lambda_{\da-} - \lambda_{\A+}\C^i_{\A\da} [\phi^i, \lambda_{\da-} ] \bigg),
\fe
where $D_\mu \equiv \partial_\mu - i [A_\mu, \cdot]$ is the gauge-covariant derivative in the adjoint representation,\footnote{Our convention is such that $x^\mu = (x^0, x^1)$, and $D_\pm \equiv {1\over 2} (D_1 \pm D_0)$.} and the gauge coupling $g_{\rm YM}$ is identified as
\ie
g_{\rm YM}^2 = { g_B\over 2\pi \ell_B^2}.
\fe
Applying the S-duality transformation to (\ref{dsaiibdoness}) yields the purely (NS,NS) spacetime background\footnote{Note that the 3-form field strength $H_3 = dB_2 = e^{2\Phi} {6dr\over r}\wedge dt\wedge dx$ obeys $\wt g_B^{-2}\int_{S^7} e^{-2\Phi} * H = (2\pi \sqrt{\A'})^6 N$, which is the electric (NS,NS) flux sourced by $N$ fundamental strings.}
\ie{}
& d s_{\rm str}^2 =  (\wt f_1(r))^{-1} (-dt^2 + dx^2) + dr^2 + r^2 d\Omega_7^2,
\\
& e^{\Phi} = (\wt f_1(r))^{-{1\over 2}},~~~~ B_2 = \wt f_1^{-1} dt\wedge dx,
\\
& \wt f_1(r) = {c_1 N\over r^6},~~~~ c_1 = 32\pi^2 \wt g_B^{1\over 2} M_{\rm pl}^{-6} = 32\pi^2 \wt g_B^2 \wt\ell_B^6,
\fe
where $\wt g_B = g_B^{-1}$ and $\wt\ell_B = g_B^{1\over 2} \ell_B$ are the string coupling and length in the dual frame.
We now compactify $x\sim x + 2\pi R$, and perform T-duality in the $x$-circle to arrive at type IIA string theory in the compactified pp-wave spacetime
\ie\label{iiappcompa}
ds_{\rm str}^2 = 2dtd\wt x + {c_1 N\over r^6} d\wt x^2 + dr^2 + r^2 d\Omega_7^2,
\fe
with vanishing dilaton $\Phi$, and the identification $\wt x \sim \wt x + 2\pi \ell_A^2/R$. The type IIA string length $\ell_A$ and coupling $g_A$ are related by
\ie
\ell_A = \wt\ell_B, ~~~~ g_A = \wt g_B {\wt\ell_B\over R} ,~~~~ c_1 = 32\pi^2 g_A^2 \ell_A^4 R^2.
\fe
We thus conclude that the type IIA string theory in background (\ref{iiappcompa}) is holographically dual to the 2D ${\cal N}=(8,8)$ $U(N)$ SYM on a circle of radius $R$, whose gauge coupling $g_{\rm YM}$ is related by
\ie\label{gagymcorr}
g_A = {1\over \sqrt{2\pi} g_{\rm YM} R}.
\fe
The MST conjecture concerns the strict $N\to \infty$ limit of the above holographic duality, where the S-matrix of type IIA string theory in asymptotically Minkowskian spacetime is proposed to be equivalent to that of the circle-compactified 2D $(8,8)$ SYM subject to the identification (\ref{gagymcorr}). In the regime of large $g_{\rm YM} R$, the gauge theory may be described in terms of an irrelevant deformation of the superconformal symmetric product orbifold ${\rm Sym}^N(\mathbb{R}^8)$, whose conformal perturbation theory is naturally identified with the type IIA superstring perturbation theory in the lightcone gauge \cite{Arutyunov:1997gi, Dijkgraaf:2003nw, mstrevisited}.

The MST dictionary is naturally extended to the non-perturbative S-matrix of type IIA string theory, whose asymptotic states include D0-branes and their bound states. These admit descriptions in the dual gauge theory as flux sector states, as will be elaborated in the next section.

\section{D0-branes as flux states}

The 2D $(8,8)$ $U(N)$ SYM admits a $\mathbb{Z}_N$ center 1-form symmetry, under which a Wilson line operator in the tensor product of $k$ fundamental representations carries charge $k$ mod $N$. The full Hilbert space of the 2D SYM compactified on a circle splits into the direct sum of superselection sectors ${\cal H}_k$, where $k$ labels the number of units of electric flux (mod $N$) \cite{Witten:1995im}. In the MST duality, where $N$ is taken to infinity, the flux sector states with finite $k$ are mapped to states in the type IIA string theory that carry $k$ units of D0-brane charge \cite{Dijkgraaf:1997vv}.

Indeed, the ground state energy of the $k$-flux sector is\footnote{(\ref{eformaf}) can also be obtained from the difference between the $(m,n)$ and $(0,n)$ string tensions \cite{Witten:1995im} in the decoupling limit.}
\ie\label{eformaf}
E_k =  {k^2 g_{\rm YM}^2\over 2 N} 2\pi R =  {\A' \over 2 N R} (kM_{\rm D0})^2,
\fe
where $M_{\rm D0} = {1\over g_A \sqrt{\A'}}$ is the mass of a single D0-brane. (\ref{eformaf}) is in agreement with the expected dispersion relation of the BPS bound state of $k$ D0-branes with vanishing transverse momentum, whose lightcone momenta are identified as $p^+ = {NR\over \A'}$ and $p^- = E_k$. 

It was pointed out in \cite{Kologlu:2016aev} that the $k$-flux sector of the 2D $(8,8)$ $U(N)$ SYM in infinite space ($R=\infty$) admits an IR description as the superconformal symmetric product orbifold ${\rm Sym}^D(\mathbb{R}^8)$, with $D = {\rm gcd}(k,N)$. The latter is in agreement with the expected transverse motion of $D$ bound particles of D0-branes, each carrying charge $k/D$ and lightcone momentum $p^+ = {NR\over D\A'}$.


We will now focus on the $k=1$ sector. The ground state energy as well as the massless degrees of freedom are accounted for by the diagonal Abelian gauge multiplet that decouples. The remaining $SU(N)$ SYM in its $k=1$ flux sector has a supersymmetric and gapped vacuum state. The MST duality predicts that in the large $N$ limit, the flux sector has massive excitations that are described by the Fock space states of massive open string modes on a D0-brane. The excitation energy is given by 
\ie
\Delta E &= {\A' \over 2 N R} \left[ \left(M_{\rm D0} + {\sum_i n_i \sqrt{\ell_i} \over \sqrt{\A'}}\right)^2 - M_{\rm D0}^2\right]
\\
&= {\sqrt{2\pi} g_{\rm YM} \over N} \left[\sum_i n_i \sqrt{\ell_i} + {\cal O}\left({1\over g_{\rm YM} R}\right)\right] ,
\label{excitationEnergy}
\fe
where $n_i \in \mathbb{Z}_{\geq 0}$ is the occupation number and $\ell_i\in \mathbb{Z}_{\geq 1}$ is the open string oscillator level. Taking the $R\to \infty$ limit, this suggests that the massive open string modes on a D0-brane are in correspondence with the massive particles of the $SU(N)$ SYM in the $k=1$ flux vacuum, whose mass values are
\ie\label{mias}
m_i = {g_{\rm YM}\over N} \sqrt{2\pi \ell_i}.
\fe
As an example, the level $\ell=1$ particles are in correspondence with open string states of mass $1/\sqrt{\A'}$ on a D0-brane. The latter come in multiplets with respect to the 16 supersymmetries preserved by the D0-brane, whose bosonic components are represented in the lightcone gauge of the worldsheet theory by the states\footnote{In the covariant quantization of the open strings (see e.g. \cite{Polchinski:1998rr}), the states (\ref{lcopens}) are represented by the worldsheet matter SCFT vertex operators $
\left( {1\over 6} d_{\mu\nu\rho} \psi^\mu \psi^\nu \psi^\rho + i e_{\mu\nu} \psi^\mu \partial X^\nu + f_\mu \partial\psi^\mu \right)  e^{i (k_+ X^+ + k_- X^-)}$,
where $X^+$ and $X^-$ are constrained by the boundary condition corresponding to the D0-brane worldline, along which the transverse coordinates $X^i$ ($i=1,\cdots,8$) take constant values, and the lightcone momenta $k_+,k_-$ obey the mass-shell condition $2 k_+ k_- = {1\over \A'}$. $d_{[\mu\nu\rho]}, e_{\mu\nu}, f_\mu$ are polarization tensors subject to the constraints $d_{\mu\nu\rho} k^\rho - e_{[\mu\nu]} =  \A' e_{\mu\nu} k^\mu +  f_\nu= e^\mu{}_\mu + 2 f\cdot k = 0$,
and the gauge redundancy $
d_{\mu\nu\rho} \sim d_{\mu\nu\rho}  - 3{\A'} \zeta_{[\mu\nu} k_{\rho]}$, $e_{\mu\nu}\sim e_{\mu\nu} + \zeta_{[\mu\nu]} + 2 k_\mu\varepsilon_\nu + \eta_{\mu\nu} \theta$, $f_\mu \sim f_\mu + \varepsilon_\mu + \A' k_\mu \theta$
for any $\zeta_{[\mu\nu]}, \varepsilon, \theta$.
These conditions can be solved in terms of the unconstrained transverse components $d_{[ijk]}, e_{ij}, f_i$. Note that these open string states assemble into the rank-3 anti-symmetric tensor and rank-2 symmetric traceless tensor of the $SO(9)$ that rotates the 9 dimensions transverse to the D0-brane worldline. However, this $SO(9)$ symmetry is not manifest in the 2D SYM at finite $N$.}
\ie\label{lcopens}
\psi_{-{1\over 2}}^i \psi_{-{1\over 2}}^j \psi_{-{1\over 2}}^k |0\rangle,~~~~ \psi_{-{1\over 2}}^i \alpha_{-1}^j |0\rangle,~~~~ \psi_{-{3\over 2}}^i |0\rangle,
\fe
where $i,j,k=1,\cdots,8$ are vector indices with respect to the $SO(8)$ rotation of the transverse directions to the lightcone coordinates, $\A_n^i$ and $\psi_r^i$ are the worldsheet transverse boson and fermion oscillators, and $|0\rangle$ stands for the oscillator ground state. Correspondingly, the lightest massive particles in the flux vacuum of the $SU(N)$ SYM, of mass $m = \sqrt{2\pi} g_{\rm YM}/N$, come in supermultiplets whose bosonic components transform in the vector, (unconstrained) rank-2 tensor, and the rank-3 anti-symmetric tensor representations of the $SO(8)$ R-symmetry respectively.


Let us note that while a priori the massive particles also give rise to states that carry nonzero spatial momenta, the latter come with excitation energy at least of order $1/R$, which is much greater than the scale $g_{\rm YM}/N$ that characterizes excitations of the D0-brane and hence decouple in the MST limit.
Furthermore, while (\ref{mias}) is a prediction of the mass spectrum in the flux vacuum in the infinite $N$ limit, we expect that at finite but parametrically large $N$ all but the first few ($\ell\leq 3$) massive states are metastable.


Let us further note that the fundamental Wilson line may be viewed as a domain wall that interpolates between the gapless non-flux vacuum and the gapped $k=1$ flux vacuum. In the infrared limit, the Wilson line should flow to a conformal boundary condition of the symmetric product orbifold SCFT ${\rm Sym}^N(\mathbb{R}^8)$. A natural guess is that this conformal boundary condition in the infinite $N$ limit is inherited from the D0-brane boundary condition of the worldsheet theory of type IIA superstring in the lightcone gauge.


\section{Resonances and their semiclassical description}

The massive open string modes on a D0-brane give rise to excited D0-brane resonances that can decay by emitting closed strings. Consider a D0-brane with a single massive open string excitation say at oscillator level $\ell=1$, whose invariant mass is $M = M_{\rm D0} + {1\over \sqrt{\A'}}$. Given the lightcone momentum $p^+ = {NR\over \A'}$, we have $p^- = {\A'\over 2N R} M^2$. The decay into the BPS D0-brane with the emission of a closed string that carries lightcone momentum $p'^+ = {nR\over \A'}$ is kinematically forbidden unless
\ie\label{kinemacon}
{\A' \over 2NR} M^2 > {\A'\over 2(N-n) R} M_{\rm D0}^2.
\fe
At weak string coupling $g_A$, and assuming $n\ll N$ (which holds in the least kinematically constrained decay channels), the condition (\ref{kinemacon}) amounts to $2g_A N/n > 1$, or in terms of the gauge theory parameters,
\ie\label{decaycond}
{n \sqrt{2\pi} g_{\rm YM} R \over 2N} < 1.
\fe
In particular, such decay is forbidden in infinite space i.e. the $R\to \infty$ limit.

Indeed, from the gauge theory perspective, the emission of a closed string by the D0-brane corresponds to a transition into the $k=1$ flux sector of the $U(N-n)$ gauge theory, whose ground state energy density is ${g_{\rm YM}^2\over 2(N-n)}$. When (\ref{decaycond}) is satisfied, a massive excitation of energy $g_{\rm YM}/N$ (as in (\ref{mias})) can decay into the $U(N-n)$ flux state, with the excess energy carried away by a low-energy excitation of the remaining $U(n)$ gauge theory. The open-closed string perturbation theory predicts the decay width of the gauge theory resonances on a circle of radius $R$ to be of order $\Gamma \sim {1\over NR}$.\footnote{This can be calculated to leading order in the expansion in $g_A$ via the disc diagram with one open and one closed string vertex operator insertion, similarly to the computation performed in \cite{Hashimoto:1996kf}.}

In the $SU(N)$ gauge theory description, it is tempting to consider a sort of Coulomb branch effective theory of the $k=1$ flux sector by expanding around the field configuration
\ie
\phi^i = \begin{pmatrix} -{x^i\over N-n} \mathbb{I}_{N-n} & 0 \\ 0 & {x^i\over n} \mathbb{I}_n \end{pmatrix},
\fe
where the gauge group is broken to $SU(N-n)\times SU(n)\times U(1)$. At large $|x^i|$, the lowest energy configuration is such that the $SU(N-n)$ gauge theory is in its $k=1$ flux vacuum, and there is a residual $U(1)$ flux that gives rise to the energy density
\ie
V_\infty = {g_{\rm YM}^2 n\over 2 N(N-n)} .
\fe
In the $n=1$ case, one expects in the large $|x^i|$ regime a massless effective field theory of the long wavelength fluctuation modes of $x^i$ and its fermionic partner. The chain of dualities in section \ref{sec:mstsetup} suggests that this effective field theory may be identified with that of a fundamental type IIA string in the lightcone gauge. In particular, the configuration where $x^i$ takes large value on an interval of the spatial circle may be identified with an open string attached to the D0-brane; at the ends of the interval where $x^i$ approaches the origin, the massless EFT breaks down, corresponding to the open string ending on the D0-brane. This EFT should in particular capture the aforementioned massive resonances in the regime of large $SO(8)$ quantum numbers.

For instance, a folded macroscopic open string spinning around the D0-brane in the $(x^1, x^2)$-plane corresponds to a soliton in the 2D SYM described by a classical solution of  the large $|x^i|$ EFT of the form
\ie
x^1 + i x^2 = r_0 \cos(\omega \sigma) e^{i\omega t},~~~~ |\sigma|< {\pi\over 2\omega},
\fe
where $(t,\sigma)$ are the 2D Minkowskian coordinates of the gauge theory. The energy $E$ and the angular momentum $J$ of the soliton are given by
\ie{}
& E = V_\infty {\pi\over\omega} + {1\over 2 g_{\rm YM}^2} \int_{-{\pi\over 2\omega}}^{\pi\over 2\omega} d\sigma \left( (\partial_t x^i)^2 + (\partial_\sigma x^i)^2 \right) = {\pi g_{\rm YM}^2 \over 2 N^2 \omega} + {\pi \omega r_0^2 \over 2 g_{\rm YM}^2},
\\
& J = {1\over g_{\rm YM}^2} \int_{-{\pi\over 2\omega}}^{\pi\over 2\omega} d\sigma (x^1 \partial_t x^2 - x^2 \partial_t x^1)
= {\pi\over 2 g_{\rm YM}^2}  r_0^2.
\fe
Minimizing $E$ at fixed $J$ determines $r_0 = {g_{\rm YM}^2\over N \omega}$, and therefore
\ie
E = {g_{\rm YM}\over N} \sqrt{2\pi J},
\fe
in agreement with the leading Regge trajectory of the spectrum (\ref{mias}).

A similar semiclassical analysis can be performed for the soliton in the $k>1$ flux sector, with the replacement $(V_\infty, r_0, E, J) \to (k^2 V_\infty, k r_0, k^2 E, k^2 J)$, leading to the result $E = k \frac{g_{\rm YM}}{N} \sqrt{2\pi J}$. This is in agreement with the leading Regge trajectory of massive open strings on $k$ D0-branes (that may be either bound or co-moving), as one would obtain by replacing $M_{D0} \to k M_{D0}$ in (\ref{excitationEnergy}).

\bigskip

In conclusion, the non-perturbative matrix string theory duality makes striking predictions on the spectrum of massive particles in the flux vacuum of the 2D $(8,8)$ $SU(N)$ SYM in the large $N$ limit, as well as their decay on a spatial circle of finite radius. A direct verification of these properties of the gauge theory, which may in principle be accessible through Hamiltonian truncation/DLCQ \cite{Antonuccio:1998tm, Dempsey:2022uie, Dempsey:2023fvm, Dempsey:2024alw} or matrix quantum mechanics bootstrap methods \cite{Lin:2023owt, Lin:2024vvg}, would either serve as highly nontrivial non-perturbative tests of the matrix string theory duality or disprove the latter.


\section*{Acknowledgements}

We are grateful to Silviu Pufu for discussions. The work of MC is supported by Clay C\'ordova's Sloan Research Fellowship from the Sloan Foundation. BG is supported by Simons Foundation grant \#994310
(Simons Collaboration on Confinement and QCD Strings). XY thanks ICTP-SAIFR, S\~ao Paulo, Brazil, the Yukawa Institute for Theoretical Physics at Kyoto University, and the organizers of the workshop “Progress of Theoretical Bootstrap” for their hospitality during the course of this work. This work is supported by DOE grant DE-SC0007870.

\bibliographystyle{JHEP}
\bibliography{MSTrefs}

@article{Hashimoto:1996kf,
	archiveprefix = {arXiv},
	author = {Hashimoto, A. and Klebanov, Igor R.},
	doi = {10.1016/0370-2693(96)00621-1},
	eprint = {hep-th/9604065},
	journal = {Phys. Lett. B},
	pages = {437--445},
	reportnumber = {PUPT-1612},
	title = {{Decay of excited D-branes}},
	volume = {381},
	year = {1996}}

@article{Sen:1997we,
	archiveprefix = {arXiv},
	author = {Sen, Ashoke},
	doi = {10.4310/ATMP.1998.v2.n1.a2},
	eprint = {hep-th/9709220},
	journal = {Adv. Theor. Math. Phys.},
	pages = {51--59},
	reportnumber = {MRI-PHY-P97-0926},
	title = {{D0-branes on T**n and matrix theory}},
	volume = {2},
	year = {1998}}

@article{Seiberg:1997ad,
	archiveprefix = {arXiv},
	author = {Seiberg, Nathan},
	doi = {10.1103/PhysRevLett.79.3577},
	eprint = {hep-th/9710009},
	journal = {Phys. Rev. Lett.},
	pages = {3577--3580},
	reportnumber = {IASSNS-HEP-97-108},
	title = {{Why is the matrix model correct?}},
	volume = {79},
	year = {1997}}

@article{Dempsey:2023fvm,
	archiveprefix = {arXiv},
	author = {Dempsey, Ross and Klebanov, Igor R. and Pufu, Silviu S. and S{\o}gaard, Benjamin T.},
	doi = {10.1007/JHEP08(2024)009},
	eprint = {2311.09334},
	journal = {JHEP},
	pages = {009},
	primaryclass = {hep-th},
	reportnumber = {PUPT-2649},
	title = {{Lattice Hamiltonian for adjoint QCD$_{2}$}},
	volume = {08},
	year = {2024}}

@article{Dempsey:2022uie,
	archiveprefix = {arXiv},
	author = {Dempsey, Ross and Klebanov, Igor R. and Lin, Loki L. and Pufu, Silviu S.},
	doi = {10.1007/JHEP04(2023)107},
	eprint = {2210.10895},
	journal = {JHEP},
	pages = {107},
	primaryclass = {hep-th},
	reportnumber = {PUPT-2635},
	title = {{Adjoint Majorana QCD$_{2}$ at finite N}},
	volume = {04},
	year = {2023}}

@article{Dempsey:2024alw,
	archiveprefix = {arXiv},
	author = {Dempsey, Ross and Pufu, Silviu S. and S{\o}gaard, Benjamin T. and Klebanov, Igor R.},
	doi = {10.1007/JHEP06(2025)260},
	eprint = {2409.19164},
	journal = {JHEP},
	pages = {260},
	primaryclass = {hep-th},
	reportnumber = {PUPT-2654},
	title = {{More about the lattice Hamiltonian for Adjoint QCD$_{2}$}},
	volume = {06},
	year = {2025}}

@article{Antonuccio:1998tm,
	archiveprefix = {arXiv},
	author = {Antonuccio, F. and Lunin, O. and Pinsky, S. and Pauli, H. C. and Tsujimaru, S.},
	doi = {10.1103/PhysRevD.58.105024},
	eprint = {hep-th/9806133},
	journal = {Phys. Rev. D},
	pages = {105024},
	reportnumber = {OHSTPY-HEP-T-98-011},
	title = {{The DLCQ spectrum of N=(8,8) superYang-Mills}},
	volume = {58},
	year = {1998}}

@book{Polchinski:1998rr,
	author = {Polchinski, J.},
	doi = {10.1017/CBO9780511618123},
	isbn = {978-0-511-25228-0, 978-0-521-63304-8, 978-0-521-67228-3},
	month = {12},
	publisher = {Cambridge University Press},
	series = {Cambridge Monographs on Mathematical Physics},
	title = {{String theory. Vol. 2: Superstring theory and beyond}},
	year = {2007}}

@article{Witten:1995im,
	archiveprefix = {arXiv},
	author = {Witten, Edward},
	doi = {10.1016/0550-3213(95)00610-9},
	eprint = {hep-th/9510135},
	journal = {Nucl. Phys. B},
	pages = {335--350},
	reportnumber = {IASSNS-HEP-95-83},
	title = {{Bound states of strings and p-branes}},
	volume = {460},
	year = {1996}}

@article{mstrevisited,
	author = {Cho, Minjae and Gabai, Barak and Scheinpflug, Jaroslav and Yin, Xi},
	journal = {in preparation},
	title = {Matrix String Theory Revisited}}

@article{Kologlu:2016aev,
	archiveprefix = {arXiv},
	author = {Kolo{\u{g}}lu, Murat},
	doi = {10.1007/JHEP11(2017)140},
	eprint = {1609.08232},
	journal = {JHEP},
	pages = {140},
	primaryclass = {hep-th},
	reportnumber = {CALT-TH-2016-028},
	title = {{Quantum Vacua of 2d Maximally Supersymmetric Yang-Mills Theory}},
	volume = {11},
	year = {2017}}

@article{Banks:1996vh,
	archiveprefix = {arXiv},
	author = {Banks, Tom and Fischler, W. and Shenker, S. H. and Susskind, Leonard},
	doi = {10.1201/9781482268737-37},
	eprint = {hep-th/9610043},
	journal = {Phys. Rev. D},
	pages = {5112--5128},
	reportnumber = {RU-96-95, SU-ITP-96-12, UTTG-13-96},
	title = {{M theory as a matrix model: A conjecture}},
	volume = {55},
	year = {1997}}

@unpublished{MaldaTalk,
	addendum = {Strings2024},
	author = {Juan Maldacena},
	note = {\url{https://indico.cern.ch/event/1284995/contributions/5975486/}},
	title =	{{The BFSS conjecture, a review}},
	year = 2024}

@article{Lin:2023owt,
	archiveprefix = {arXiv},
	author = {Lin, Henry W.},
	doi = {10.1007/JHEP06(2023)038},
	eprint = {2302.04416},
	journal = {JHEP},
	pages = {038},
	primaryclass = {hep-th},
	title = {{Bootstrap bounds on D0-brane quantum mechanics}},
	volume = {06},
	year = {2023}}

@article{Lin:2024vvg,
	archiveprefix = {arXiv},
	author = {Lin, Henry W. and Zheng, Zechuan},
	doi = {10.1007/JHEP01(2025)190},
	eprint = {2410.14647},
	journal = {JHEP},
	pages = {190},
	primaryclass = {hep-th},
	title = {{Bootstrapping ground state correlators in matrix theory. Part I}},
	volume = {01},
	year = {2025}}

@article{Motl:1997th,
	archiveprefix = {arXiv},
	author = {Motl, Lubos},
	eprint = {hep-th/9701025},
	month = {1},
	reportnumber = {HEP-UK-0003},
	title = {{Proposals on nonperturbative superstring interactions}},
	year = {1997}}

@article{Banks:1996my,
	archiveprefix = {arXiv},
	author = {Banks, Tom and Seiberg, Nathan},
	doi = {10.1016/S0550-3213(97)00278-2},
	eprint = {hep-th/9702187},
	journal = {Nucl. Phys. B},
	pages = {41--55},
	reportnumber = {RU-97-06},
	title = {{Strings from matrices}},
	volume = {497},
	year = {1997}}

@article{Dijkgraaf:1997vv,
	archiveprefix = {arXiv},
	author = {Dijkgraaf, Robbert and Verlinde, Erik P. and Verlinde, Herman L.},
	doi = {10.1016/S0550-3213(97)00326-X},
	eprint = {hep-th/9703030},
	journal = {Nucl. Phys. B},
	pages = {43--61},
	reportnumber = {CERN-TH-97-034, CERN-TH-97-34, THU-97-06, UTFA-97-06},
	title = {{Matrix string theory}},
	volume = {500},
	year = {1997}}

@article{Arutyunov:1997gi,
	archiveprefix = {arXiv},
	author = {Arutyunov, G. E. and Frolov, S. A.},
	doi = {10.1016/S0550-3213(98)00326-5},
	eprint = {hep-th/9712061},
	journal = {Nucl. Phys. B},
	pages = {159--206},
	title = {{Four graviton scattering amplitude from S**N R**8 supersymmetric orbifold sigma model}},
	volume = {524},
	year = {1998}}

@article{Dijkgraaf:2003nw,
	archiveprefix = {arXiv},
	author = {Dijkgraaf, Robbert and Motl, Lubos},
	eprint = {hep-th/0309238},
	month = {9},
	reportnumber = {HEP-UK-0019, HUTP-03-A063, ITFA-2003-45},
	title = {{Matrix string theory, contact terms, and superstring field theory}},
	year = {2003}}

\end{document}